\documentclass[reprint,a4paper,twocolumn,aps,amsmath,amssymb,floatfix,
  superscriptaddress,showkeys,longbibliography]{revtex4-2}

\usepackage[utf8]{inputenc}
\usepackage[T1]{fontenc}
\usepackage{microtype}

\usepackage{mathtools}

\usepackage{graphicx}
\graphicspath{{./}{../}}
\usepackage[dvipsnames,svgnames,table]{xcolor}

\usepackage{siunitx}
\usepackage{enumitem}
\usepackage{comment}
\usepackage{placeins}
\usepackage{soul}
\usepackage{xspace}
\usepackage{xifthen}

\usepackage{physics}

\usepackage[colorlinks,linkcolor=blue,citecolor=blue,urlcolor=blue]{hyperref}

\def\equationautorefname~#1\null{Eq.~(#1)\null}

\definecolor{goodred}{rgb}{0.7,0,0}
\definecolor{cadet}{rgb}{0.33, 0.41, 0.47}
\definecolor{plum(traditional)}{rgb}{0.56, 0.27, 0.52}

\definecolor{blue-violet}{rgb}{0.54, 0.17, 0.89}

\begin{document}

\title{Quantum Dynamical Signatures of Topological Flow Transitions in Limit Cycle Phases}
\author{Alejandro S. Gómez}
\email{alejandros.gomez@uam.es}
\affiliation{Departamento de Física Teórica de la Materia Condensada and Condensed Matter Physics Center (IFIMAC),
Universidad Autónoma de Madrid, E-28049 Madrid, Spain}
\author{Javier del Pino}
\email{j.delpino@uam.es}
\affiliation{Departamento de Física Teórica de la Materia Condensada and Condensed Matter Physics Center (IFIMAC),
Universidad Autónoma de Madrid, E-28049 Madrid, Spain}

\begin{abstract}
Quantum self-oscillatory phases are ubiquitous in driven-dissipative systems. Classically, each phase is defined by its flow pattern and how stationary sets organize phase space (e.g., fixed points and limit cycles), with transitions triggered by local bifurcations or global basin rearrangements.
In the quantum regime, these reorganizations are often blurred by density-matrix averaging, and spectral indicators such as the Liouvillian gap can miss changes that unfold mainly in the transients.
Here, we develop a topological toolkit for flows with limit cycles, rooted in Morse-Smale dynamical-systems theory, to reveal and classify dynamical phase transitions in self-oscillating driven-dissipative systems. We formulate a graph invariant, the \textit{molecule}, encoding the connectivity and chirality of fixed points and limit cycles and the resulting \textit{topological} constraints on basin and flow-region rearrangements during phase transitions.
Exploiting these constraints, we uncover a class of global dynamical phase transitions, marked by discrete changes of the molecule, that leave no clear trace in the Liouvillian spectrum (gap or excited eigenvalues) yet induce sharp qualitative changes in transient quantum dynamics and relaxation pathways.
Our findings show that flow topology offers a clear and unified way to identify and classify dynamical phases beyond what Liouvillian spectra alone reveal.
\end{abstract}
\maketitle

\section{Introduction}
Driven-dissipative quantum systems provide a rich arena to explore non-equilibrium dynamical phases of matter. Of particular interest are self-oscillatory states, where the system sustains oscillations at frequencies distinct from the drive. They appear as subharmonic responses (e.g., period-doubling in parametrically driven resonators)~\cite{leuch_parametric_2016,kovacic_mathieus_2018,grimm_kerr-cat_2020,eichler2023classical} and as limit cycles (LCs), isolated orbits in phase space with spontaneous oscillation frequencies set by the balance of nonlinearity, dissipation, and coherent drive~\cite{Keeling2010,Lee2013,lorch_laser_2014,ben_arosh_quantum_2021,kongkhambut_observation_2022,kehrer_2025,liu_2025,Seibold2020}. Driven-dissipative phases arise naturally in quantum platforms engineering parametric amplification and strong nonlinear interactions, such as superconducting devices, cold atoms, trapped ions, and nanomechanical systems~\cite{ripoll2022quantum,ritsch_cold_2013,leibfried_quantum_2003,poot_mechanical_2012}. Moreover, leveraging such states is key to quantum control~\cite{brif_control_2010}, generating frequency combs~\cite{cundiff_2003}, stabilizing error-correction codes~\cite{cai_bosonic_2021}, and realizing collective many-body phases~\cite{hartmann_quantum_2008,eisert2015quantum,Abanin2019}.

Self-oscillations appear stationary in a suitable rotating frame, allowing dynamical phases to be classified using intuition from time-independent systems~\cite{soriente_distinctive_2021}. In this picture, phase transitions occur when a quasienergy gap of the Floquet Hamiltonian closes~\cite{oka2019floquet}. In open driven systems, the same logic applies to the Liouvillian, which evolves the density matrix: closing the real part of its spectrum marks a dissipative transition, while closing its complex (non-Hermitian) gap can reorganize the oscillatory modes of the system~\cite{minganti_spectral_2018,haga_oscillating-mode_2024}. However, these spectral indicators are insufficient to determine the specific dynamical change, e.g., local versus global, because the density matrix averages over phase space. This limitation is critical across parametric instabilities or superradiant transitions, which generate coexisting self-oscillatory responses (e.g., period-doubling and LCs) mapped to distinct phase-space features
~\cite{Chiacchio2023,delpino2024,fu_sideband_2025}.

In the large-excitation limit, the dynamics reduce to a phase-space flow organizing all deterministic trajectories. Each distinct flow pattern then defines a \textit{phase}, whose long-term behavior is set by its attractors, typically fixed points (FPs) or LCs. In 2D (Morse) flows
containing only FPs, a graph invariant captures their persistent qualitative pattern under small parameter changes.~\cite{oshemkov_classification_1998} In driven-dissipative systems, FP attractors also carry a symplectic norm distinguishing particle- and hole-like behavior~\cite{blaizot1986quantum,Rossignoli2005,flynn_deconstructing_2020,soriente_distinctive_2021}, which must enter the invariant~\cite{villa2025topological}. The invariant changes only under flow reorganizations, from local FP stability or symplectic-norm flips, captured by linear response, ~\cite{soriente_distinctive_2021,Dumont2024,seibold_2025} to global bifurcations away from FPs.
But once LCs enter, reorganizations of LCs and FPs become intertwined, making the FP invariant incomplete. For instance, LCs set extended phase-space boundaries that constrain local changes and also undergo their own global events, such as saddle-loops or LC-LC collisions.~\cite{glendinning1994stability,guckenheimer2013nonlinear} A key question follows: how do we build an invariant that incorporates both FPs and LCs, captures their global transitions, and predicts their physical signatures?

In this work, we build on the topological classification of driven-dissipative FP phases established in Ref.~\cite{villa2025topological} and extend it to self-oscillating phases with LCs. By extending the corresponding topological invariant, we uncover a family of phase transitions with unique quantum dynamical signatures. 
Guided by the classification of \textit{structurally stable} (qualitatively robust) flows with LCs, called Morse-Smale flows~\cite{oshemkov_classification_1998}, we define a \textit{molecule}, an invariant encoding the arrangement and chirality of FPs and LCs on equal footing, with FP chirality set by the symplectic norm in the high-excitation limit.
The molecule condenses the full deterministic phase-space dynamics into a single visual object. Its structure, set by the orientation and ordering of neighboring flow regions, dictates how the invariant can change at bifurcations and thus which phase transitions are topologically allowed or forbidden.
Because the flow provides the scaffold for the quantum dynamics, these transitions imprint clear signatures in transient quantum states, even though they need not clearly manifest in the Liouvillian spectrum.
As an illustration, we apply the framework to a parametrically driven Kerr resonator with nonlinear loss and gain, a model capturing the core mechanisms of self-oscillatory phases across quantum platforms such as trapped ions, levitated systems, light–matter ensembles, superconducting circuits, and hybrid quantum devices.

\section{Reference model}

\begin{figure}[t]
  \centering
  \includegraphics[width=\linewidth]{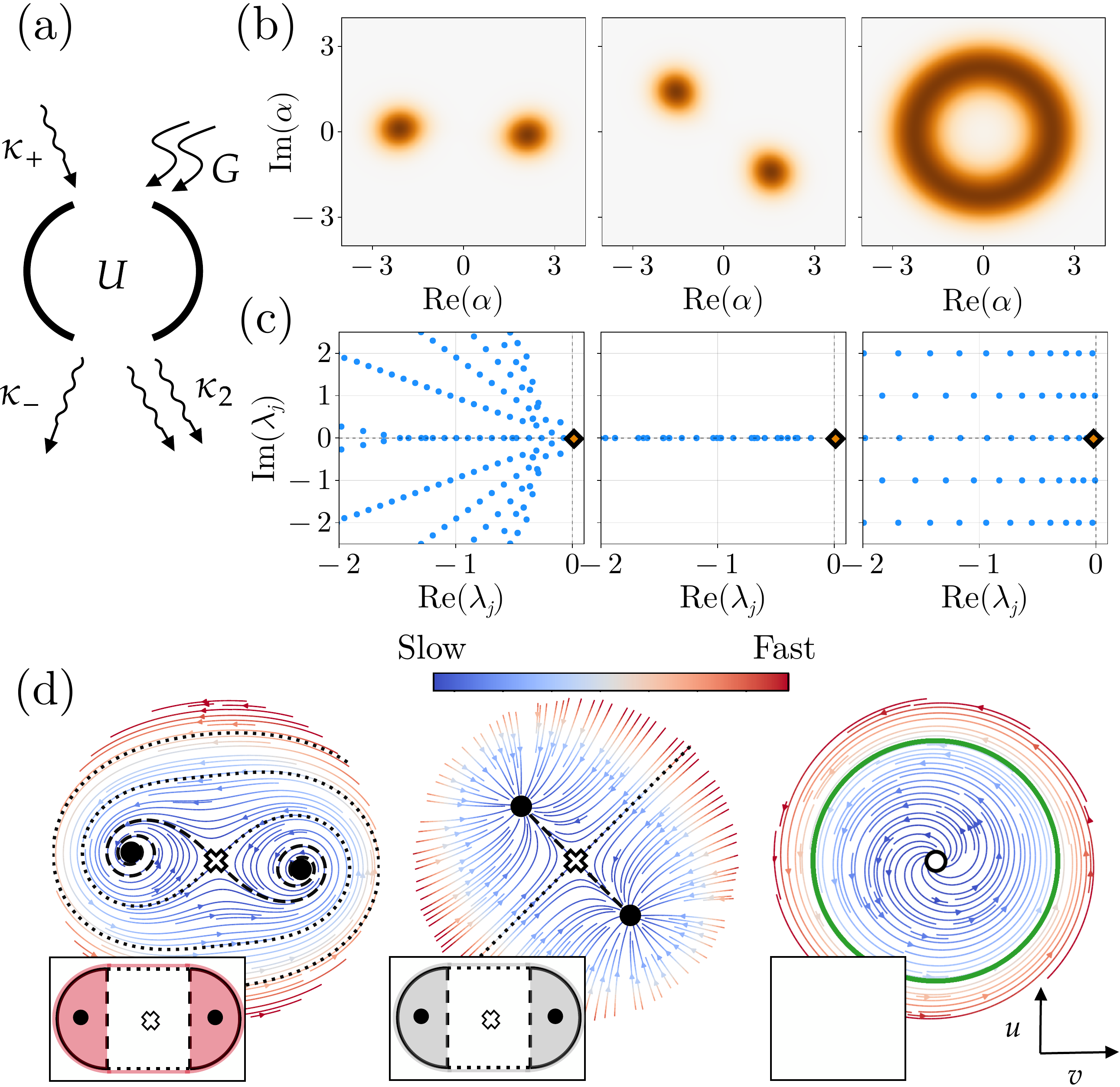}
  \caption{\textit{Flow topology of driven-dissipative nonlinear resonators} (a) Driven-dissipative resonator with Kerr nonlinearity $U$, two-photon drive $G$, single- and two-photon loss $\kappa_-,\kappa_2$, and single-photon gain $\kappa_+$. (b) Steady-state Wigner function for a Kerr Parametric Oscillator (KPO, $\kappa_2=0=\kappa_+$, $U=-0.1$, $G=0.25$, $\Delta=0$), a two-photon driven dissipative oscillator (2DPO, $U=0=\kappa_+$, $G=0.25$, $\kappa_2=0.1$, $\Delta=0$), and a Rayleigh Van der Pol oscillator (RVdP, $\kappa_2=0.01, \kappa_+=0.1$, $U=-0.1$, $\kappa_-=0$, $G=0$ and $\Delta=1$).
  For the KPO and 2DPO cases, the single-photon loss is set to $\kappa_-=0.1$. (c) Corresponding Liouvillian spectrum and (d) mean-field vector flows in the rotating quadratures $(u,v)$. FP attractors (repellers), LCs, and saddles appear as black (white) dots, green curves, and white crosses, respectively. Boxed insets show the flow-topology graph invariants built from the FPs and their separatrices~\cite{villa2025topological}. Red (gray) shading marks regions of negative (zero) chirality in the basins of attraction of the corresponding FPs. A flow containing an LC falls outside the classification.
  }
  \label{fig:1}
\end{figure}

Our model consists of a nonlinear Kerr parametric oscillator of natural frequency $\omega_0$, with annihilation operator $\hat{a}$ driven near parametric resonance at frequency $2\omega$ and subject to linear gain and nonlinear loss [Fig.~\ref{fig:1}(a)]. In the lab frame, the system is invariant under discrete time translations $t\mapsto t+nT$ by multiples of its drive period $T=2\pi/\omega$. Moving to a rotating frame at frequency $\omega$ (doubled period $2T$) and applying the rotating-wave approximation, the density matrix evolves as $d\hat{\rho}/dt=\mathcal{L}[\hat{\rho};\boldsymbol{\eta}]$ with time-independent Liouvillian
\begin{equation}\label{eq:general_eq}
\mathcal{L}[\square;\boldsymbol{\eta}]= -i[\hat{\mathcal{H}},\square] + \sum_i\big[ c_i\square c_i^\dagger - \tfrac{1}{2}\{c_i^\dagger c_i,\square\}\big],
\end{equation}
where Hamiltonian $\hat{\mathcal{H}}$ and collapse operators $\hat{c}_i$ read, 
\begin{align}\label{Eq:liouvillian_parts}
\hat{\mathcal{H}} =& - \Delta\hat{a}^\dagger\hat{a} +
\frac{U}{2} \hat{a}^{\dagger 2}\hat{a}^2 +
\frac{G}{2} \big(\hat{a}^{\dagger 2}+ \hat{a}^2\big), \\
\hat{c}_i \in & \{\sqrt{\kappa_-}\hat{a}, \sqrt{\kappa_+}\hat{a}^\dagger, \sqrt{\kappa_2}\hat{a}^2 \}. \label{eq:collapse_operators}
\end{align}
The vector $\boldsymbol{\eta} = (\Delta,U,G,\kappa_-,\kappa_+,\kappa_2)$ collects all parameters entering the right-hand side of Eq.~\eqref{eq:general_eq}. Here $\Delta=(\omega^2-\omega_0^2)/\omega$ is a detuning from parametric resonance, $U$ is the Kerr nonlinearity, and $G$ is the two-photon drive amplitude. The collapse operators account for single-photon loss (rate $\kappa_-$), single-photon gain (rate $\kappa_+$), and two-photon loss (rate $\kappa_2$), respectively.

\begin{figure*}[ht]
  \centering  \includegraphics[width=\textwidth]{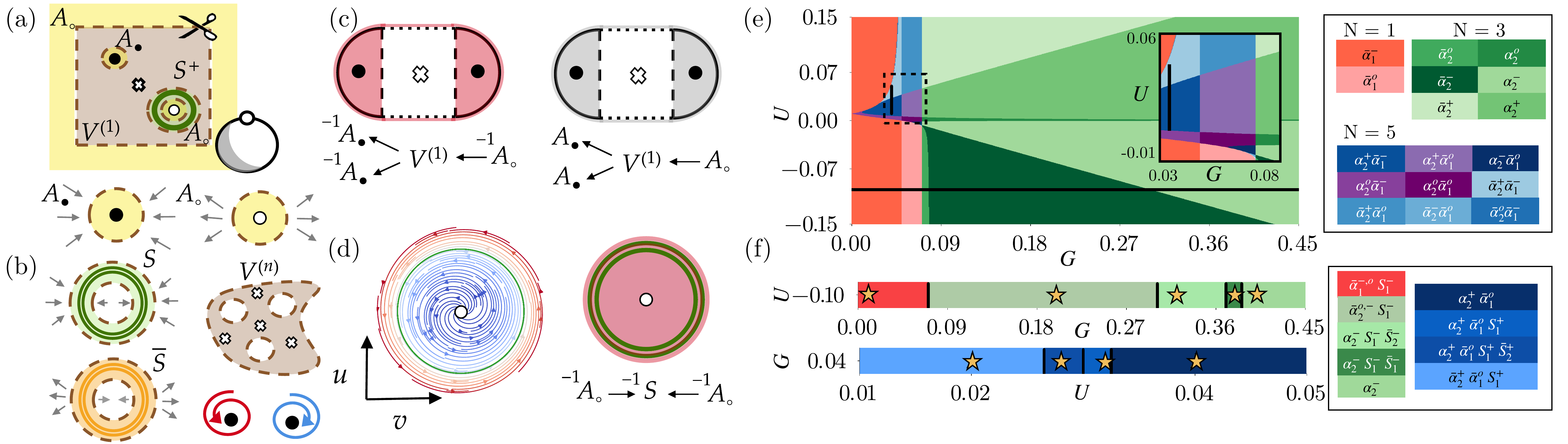}
  \caption{\textit{Fixed-point phase topology and Morse-Smale invariants in the $(U,G)$ plane.} (a) Cutting-set construction used to compactify the flow; the white dot represents the source at infinity. (b) Elementary atoms and chirality convention: FP attractor $A_\bullet$, repeller $A_{\circ}$, attracting/repelling LC $S, \bar{S}$, and saddle region $V^{(n)}$. (c) Morse-Smale graphs for the KPO and 2DPO of Fig.~\ref{fig:1}(d) and associated molecules. (d) Example of graph invariant, similar to~\cite{villa2025topological}, and molecule built by these rules for an RVdP flow with an attracting LC. (e) Phase diagram vs $(U,G)$ with legend of FP classes; the inset enlarges the dashed region. Parameters fixed (a.u.): $\Delta=0.05$, $\kappa_+-\kappa_-=0.1$, $\kappa_2=0.01$. (f) One-dimensional cuts at $U=-0.10$ and $G=0.04$ from panel (e), showing the number of attracting (repelling) LCs in each region $S_k^i (\bar{S}_k^i)$; stars mark the $(U,G)$ points for the flows in Fig.~\ref{fig:3}.
  }
  \label{fig:2}
\end{figure*}

The model in Eq.~\eqref{eq:general_eq} and Eq.~\eqref{Eq:liouvillian_parts} descends from the Mathieu-Van der Pol-Duffing oscillator~\cite{pandey2007frequency} after the rotating-wave approximation~\cite{ben_arosh_quantum_2021}. It interpolates between standard nonlinear oscillators: the Kerr parametric oscillator (KPO) for $\kappa_{2}=0$~\cite{wielinga_quantum_1993}, the two-photon-dissipative parametric oscillator (2DPO) for $U=0$ and $\kappa_{2}>0$~\cite{mirrahimi2014dynamically}, and the Rayleigh-Van der Pol oscillator (RVdP) for $G=0$ and $\kappa_{+}>0$~\cite{ben_arosh_quantum_2021}. 
A common approach to dynamical phase transitions is to track the Liouvillian spectrum as $\boldsymbol{\eta}$ varies. The steady state $\hat{\rho}_0$ corresponds to the eigenmode with vanishing decay rate, i.e., with Liouvillian eigenvalue $\lambda_0$ such that $\mathrm{Re}(\lambda_0)=0$.
In Fig.~\ref{fig:1}(b) we show the Wigner function of $\hat{\rho}_0$ in representative limits; Fig.~\ref{fig:1}(c) displays the corresponding Liouvillian spectra. In the KPO and 2DPO limits, the Wigner function is bimodal, corresponding to a coherent-state mixture with $\mathrm{Im}(\lambda_0)=0$.
In the RVdP steady state, probability concentrates along a closed orbit corresponding to an LC. Time-translation symmetry is continuously broken, as the system oscillates with an emerging frequency $\omega^\star$, which appears as a conjugate pair of Liouvillian eigenvalues $\mathrm{Im}(\lambda_{0,\pm})=\pm \omega^\star$. In the classical limit, the periodic motion produces an infinite Liouvillian ladder at all LC harmonics, $\mathrm{Im}(\lambda_{0,n})=\pm n\omega^\star$. At low photon numbers, this ladder deforms smoothly into parabolic branches~\cite{Dutta2025}.
In all these phases, the Wigner function remains positive, indicating that the steady state $\hat{\rho}_0$ admits a classical phase-space description.

\section{Flow-topology classification}
\subsection{Classification of flows with limit cycles}
We associate with $\mathcal{L}[\hat{\rho};\boldsymbol{\eta}]$ a planar flow $d\boldsymbol{r}/dt=\mathbf{f}(\boldsymbol{r},\boldsymbol{\eta})$
in the real quadratures $\boldsymbol{r}=(u,v)^T$ defined by $\mathrm{Tr}[\hat{\rho}\hat{a}]=u+iv$, via a mean-field factorization $\langle\square\triangle\rangle\simeq\langle\square\rangle\langle\triangle\rangle$, which neglects quantum correlations. From Eq.~\eqref{eq:general_eq} and Eq.~\eqref{Eq:liouvillian_parts} we find
\begin{equation}\label{Eq:SemiclassicalEOM}
\mathbf{f}(\boldsymbol{r},\boldsymbol{\eta})
= 
\begin{pmatrix}
\left( \frac{\gamma}{2}-\kappa_2 R^2\right)u - (\Delta+G-UR^2)v \\
\left(\frac{\gamma}{2}-\kappa_2R^2 \right)v +(\Delta-G-UR^2)u
\end{pmatrix},
\end{equation}
where $R=\sqrt{u^2+v^2}$ and $\gamma\equiv\kappa_+-\kappa_-$. For further details on the semiclassical analysis, see Appendix~\ref{sec:local_trans}.
The flows corresponding to the parameter values of Fig.~\ref{fig:1}(b) are shown in Fig.~\ref{fig:1}(d). Their stationary sets ($t\rightarrow\infty$) consist of
 FPs (solutions to $\mathbf{f}=0$) of attractor, repeller or saddle type, depending on the stability, and LCs (closed isolated trajectories $\boldsymbol{r}^\star(t+T^\star)=\boldsymbol{r}^\star(t)$ fulfilling Eq.~\eqref{Eq:SemiclassicalEOM}), which act as periodic attractors or repellers. The flow also contains separatrices, trajectories that mark the boundaries between regions whose initial conditions evolve toward different long-term behaviors (basins of attraction).
Bifurcations mark abrupt reorganizations of the flow. They may be \textit{local}, creating or destroying FPs across instabilities, or \textit{global}, such as separatrix reconnections and reshaping of entire basins without changing the set of FPs. %
Figures~\ref{fig:1}(b) and \ref{fig:1}(d) show that in the semiclassical regime, the steady state $\hat{\rho}_0$ inherits the structure of the phase-space flow~\eqref{Eq:SemiclassicalEOM}: the flow determines how weak quantum fluctuations circulate around and between attracting FPs and LCs, which accumulate probability in their basins. In the strongly fluctuating regime this correspondence can break down, as noise or quantum jumps can ``wash out'' the classical structure~\cite{seibold_2025}.
Quantum dynamics is fully encoded in the evolution of the density matrix, generated by the Liouvillian. Yet this Liouvillian acts on a high-dimensional space and has an infinite spectrum, making it difficult to identify which modes organize the transient dynamics. This motivates the question: can the mean-field flow, containing all individual phase-space trajectories, track transient quantum dynamics with LCs more directly and expose the associated dynamical phases?

A natural way to address this question is to collect the essential structure of the flow (FPs, LCs, and separatrices), and use this to define a topological invariant.
For generic parameters $\boldsymbol{\eta}$, the flow~\eqref{Eq:SemiclassicalEOM} is structurally stable (finitely many FPs and LCs, no saddle connections), and admits a topological invariant, see Appendix~\ref{sec:morse_smale}.
In systems combining Hamiltonian dynamics with dissipation, the chirality of FP attractors, encoded by their symplectic norm, determines whether local excitations correspond to a harmonic or inverted oscillator~\cite{soriente_distinctive_2021,Dumont2024}; without LCs, a decorated graph incorporating chirality forms a complete topological invariant~\cite{villa2025topological}.
Inspired by Morse-Smale theory~\cite{oshemkov_classification_1998}, we extend this framework to flows with LCs via a \textit{molecule} invariant [Figs.~\ref{fig:2}(a)--\ref{fig:2}(d)], whose \textit{atoms} encode FPs, LCs, and their chirality, connected by separatrices. We detail its construction below.

\subsection{Construction of the molecule invariant}
We start from the planar phase-space flow of Eq.~\eqref{Eq:SemiclassicalEOM} and compactify it into a closed surface. We identify all points reached as $|\boldsymbol r|\to\infty$ with a single point at infinity, so that the plane is wrapped into a sphere by stereographic projection [Fig.~\ref{fig:2}(a)]. To build the topological invariant, we then decompose the resulting structurally stable flow $v$ based on its asymptotic behavior~\cite{oshemkov_classification_1998}. We define an atom as an equivalence class of flow-invariant compact regions containing exactly one recurrent set (attractor or repeller) or a connected set of saddle points. In simple terms, atoms are the robust local building blocks of the flow [Fig.~\ref{fig:2}(b)], with five types: point attractors $(A_\bullet)$, point repellers $(A_\circ)$, attracting $(S)$ and repelling $(\bar{S})$ LCs, and saddle regions $V^{(n)}$ containing $n$ saddle points.

\begin{figure*}[!t]
  \centering
   \includegraphics[width=\textwidth]{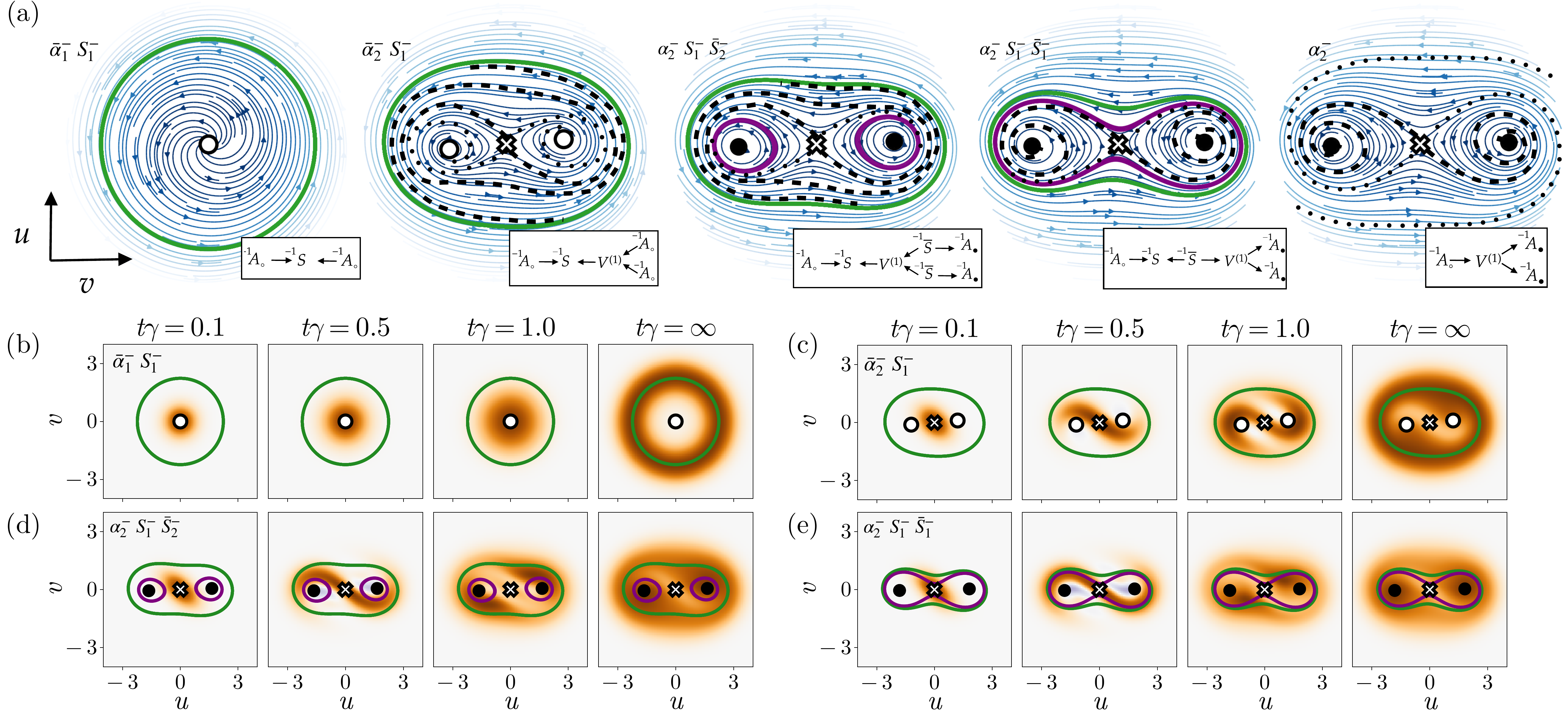}
  \caption{\textit{Flow topology and quantum evolution toward steady states.} (a) Semiclassical flows for starred points of Fig.~\ref{fig:2}(f) with $G=0.01, 0.20, 0.32, 0.38, 0.40$, (phases denoted $\bar{\alpha}_1^- S_1^-$, $\bar{\alpha}_2^-S_1^-$, $\alpha_2^- S_1^- \bar{S}_2^-$, $\alpha_2^- S_1^- \bar{S}_1^-$ and $\alpha_2^-$) together with their topological invariants (molecules). Attracting (repelling) LCs are shown as green (purple) curves, and qualitative stable (unstable) separatrices as dashed (dotted) lines. (b)--(e) Wigner function at intermediate times (see panel titles) and in the steady state ($t\to\infty$).
  }
  \label{fig:3}
\end{figure*}

To assemble these atoms into the global \textit{molecule} graph, we define a \textit{cutting set} consisting of a disjoint family of neighborhoods: a disk enclosing each FP atom $(A_\bullet,A_\circ)$, corresponding to attractors and repellers, respectively, an annulus enclosing each LC atom $(S,\bar{S})$ for attracting/repelling LCs, respectively, and a neighborhood for each saddle complex. Cutting the manifold along the boundaries of these sets isolates the atoms, which become the nodes of the graph. Then, we assign directed edges based on the separatrix flow connecting these boundaries. We distinguish three bond types, $s$, $t$, and $u$, connecting repellers to saddles, repellers to attractors, and saddles to attractors, respectively. Finally, to capture the orientation of the spiraling motion, we assign a local chirality $\sigma \in \left\{+,-,o\right\}$ to each atom defined by the local winding relative to the rotating frame frequency~\cite{soriente_distinctive_2021} CW $(+)$, CCW $(-)$ or undefined ($o$) for overdamped character, extending the classification of Morse-Smale flows~\cite{oshemkov_classification_1998}. Note that the cutting set is in general not unique, since different admissible choices of discs, annuli, or saddle neighborhoods may be used; however, all such choices lead to equivalent atom decompositions and hence to the same molecule up to graph isomorphism.

Excluding LCs, the molecule reduces to an object akin to the FP graph invariants of Ref.~\cite{villa2025topological} and Fig.~\ref{fig:1}(d); the mapping is shown in Fig.~\ref{fig:2}(c).
Conversely, the FP graph invariant extends naturally to the RVdP limit ($U=0$, $G=0$, $\gamma>0$, $\kappa_2>0$), where an attracting LC adds an $S$ vertex. This vertex shares the chirality of both the enclosing repeller and the source at infinity [Fig.~\ref{fig:2}(d)], consistent with the Poincaré-Bendixson theorem~\cite{guckenheimer2013nonlinear}.

\subsection{Limit-cycle flow topology of the driven-dissipative oscillator}

Figure~\ref{fig:2}(e) revisits the phase diagram classified by the FP invariant, which already produces a rich range of phases.
Figure~\ref{fig:2}(f) illustrates two representative cuts of this diagram, where the molecule reveals flow structure beyond the FP classification, with coexisting attracting and repelling FPs and LCs.

The flows shown in Fig.~\ref{fig:3}(a), corresponding to the starred points in Fig.~\ref{fig:2}(f), reveal the sequence of phase transitions [local (global) bifurcations labelled by $\ell$ ($g$)].
\begin{equation}
    \bar{\alpha}_1^- S_1^-\overset{\ell}{\mapsto}\bar{\alpha}_2^-S_1^-\overset{\ell}{\mapsto}{\alpha}_2^- S_1^- \bar{S}_2^-\overset{g}{\mapsto}{\alpha}_2^- S_1^- \bar{S}_1^-\overset{g}{\mapsto}{\alpha}_2^-.
\end{equation}
Starting from the RVdP phase, $\bar{\alpha}_1^- S_1^-$, with an LC and a source, the sequence proceeds as follows. (i) A pitchfork bifurcation creates a sink–saddle pair ($\varnothing \to V^{(1)}+A_\bullet$), morphing 
$\bar{\alpha}_1^- S_1^-\mapsto\bar{\alpha}_2^-S_1^-$.
(ii) A subcritical Hopf bifurcation on both repellers nucleates two unstable LCs ($A_\circ \rightarrow A_\bullet+\bar{S}^-$) and converts repellers into attractors, giving $\bar{\alpha}_2^-S_1^-\mapsto{\alpha}_2^- S_1^- \bar{S}_2^-$.
(iii) A global homoclinic saddle loop (saddle self-connection) forms when an LC expands and collides with the saddle, destroying the cycle and rewiring separatrices; by $\mathbb{Z}_2$ symmetry $(\hat{a} \to -\hat{a})$, this produces a new repelling LC immediately after the double homoclinic bifurcation ($V^{(1)} + \bar{S}_2^-  \rightarrow V^{(1)} + \bar{S}_1^-$), leading to $\alpha_2^- S_1^- \bar{S}_1^-$.
(iv) Finally, a fold bifurcation \cite{glendinning1994stability} annihilates two nested LCs of opposite stability, allowing trajectories from the former exterior basin to reach the FP attractors. 
The result is the bistable KPO phase ${\alpha}_2^-$.
When varying the Kerr nonlinearity $U$ at fixed $G$, analogous sequences arise in the five-FP region, adding more complex transitions, e.g. via global heteroclinic (saddle to saddle) connections, all captured by our invariant; see Appendix~\ref{sec:global_5FP}.

\begin{figure}[b]
     \centering
    \includegraphics[width=0.9\linewidth]{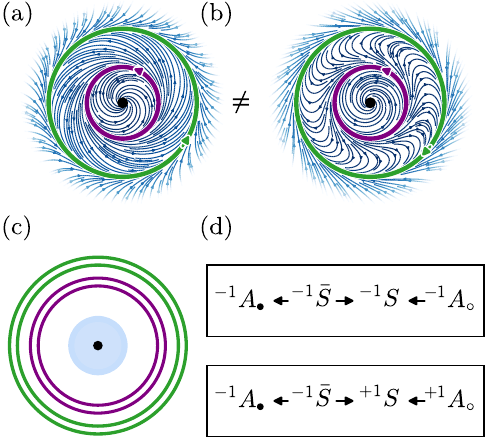}
  \caption{\textit{Geometric invariant vs graph invariant.} Comparison between two phase portraits with identical geometric invariant but different LC chiralities. Panels (a) and (b) show nested LCs with the same fixed-point and geometry, but inequivalent orientations of the inner and outer cycles. Panel (c) shows the corresponding geometric colored graph, which does not distinguish the two cases. Panel (d) shows the molecule invariant, which resolves the missing chirality information and therefore separates the two phases.}
    \label{fig:4_now}
\end{figure}

\subsection{Nested limit cycles and topological selection rules}
Although inspired by the Oshemkov-Sharko construction~\cite{oshemkov_classification_1998}, our molecule also resolves a limitation that arises when extending the decorated graph invariants of Ref.~\cite{villa2025topological} to flows with coexisting FPs and LCs. The difficulty stems from nested LC domains: a cycle may enclose, rather than simply border, other basins, making basin adjacency ambiguous in the colored graph representation. Although this representation faithfully depicts the geometric arrangement of basins on the sphere, it can obscure the topological constraints governing bifurcations, including the relative chirality between different $S$-atom regions, see Fig.~\ref{fig:4_now}.

The molecule resolves this ambiguity by using the dual graph of the phase-space cutting set, where each basin of attraction or repulsion is collapsed into a single atom, while adjacency and chirality are encoded by graph connectivity. Crucially, this makes the (topological) selection rules for allowed phase transitions explicit: only adjacent atoms, or atoms that become adjacent through a global reconnection, can participate in a local or global bifurcation.

\section{Transient signatures of flow-topology transitions}

\subsection{Hidden flow topology beyond the Liouvillian gap}

The Liouvillian spectrum is the standard probe of relaxation and dissipative phase transitions in open quantum systems. For Markovian dynamics generated by $\mathcal{L}$ (e.g. Eq.~\eqref{eq:general_eq}), each eigenvalue $\lambda_i$ sets a transient decay rate, $-\mathrm{Re}(\lambda_i)$, and oscillation frequency, $\mathrm{Im}(\lambda_i)$. The real Liouvillian gap, $\Delta_{\mathrm L}=\min_{i\neq0}[-\mathrm{Re}(\lambda_i)]$, identifies the slowest asymptotic channel; its closing signals critical slowing down and a sharp rearrangement of the long-time state~\cite{minganti_spectral_2018}. More targeted probes, such as the oscillating-mode gap, isolate the least-damped oscillatory modes and detect underdamped-to-overdamped transitions~\cite{minganti_quantum_2019,haga_oscillating-mode_2024}. Here we expose a different limitation: global reorganizations of the phase-space flow can reroute transient pathways, basins, and dynamical barriers without closing the real or oscillatory Liouvillian gaps. The molecule therefore detects flow-topological transitions that remain hidden from the low-lying Liouvillian spectrum.

In the first four phases of the sequence, $\bar{\alpha}_1^- S_1^-$, $\bar{\alpha}_2^- S_1^-$, $\alpha_2^- S_1^- \bar{S}_2^-$, and $\alpha_2^- S_1^- \bar{S}_1^-$, $\hat{\rho}_0$ contains a single attracting LC, so the Wigner distributions are essentially equivalent; only the arrangement of saddles, repellers, and repelling LCs differs.
These structures leave little trace in the steady state but shape relaxation pathways. After a quench from the vacuum [Fig.~\ref{fig:3}(b)--(e)], the fluctuations expand isotropically in $\bar{\alpha}_1^- S_1^-$ [Fig.~\ref{fig:3}(b)]. In $\bar{\alpha}_2^- S_1^-$ [Fig.~\ref{fig:3}(c)] the state is first funneled into a saddle and winds around repellers before settling on the attracting LC. In $\alpha_2^- S_1^- \bar{S}_2^-$ [Fig.~\ref{fig:3}(d)], it relaxes again to a cycle, but a repelling LC pushes it away from a finite forbidden region.
The topological framework uncovers nonlocal transitions arising from $g$ bifurcations. A clear example is the passage $\alpha_2^- S_1^- \bar{S}_2^- \to \alpha_2^- S_1^- \bar{S}_1^- \to \alpha_2^-$. The global event involves the coalescence and annihilation of LCs, marked by the expansion of a forbidden region [Fig.~\ref{fig:3}(e)].

\begin{figure}[b]
  \centering
  \includegraphics[width=\linewidth]{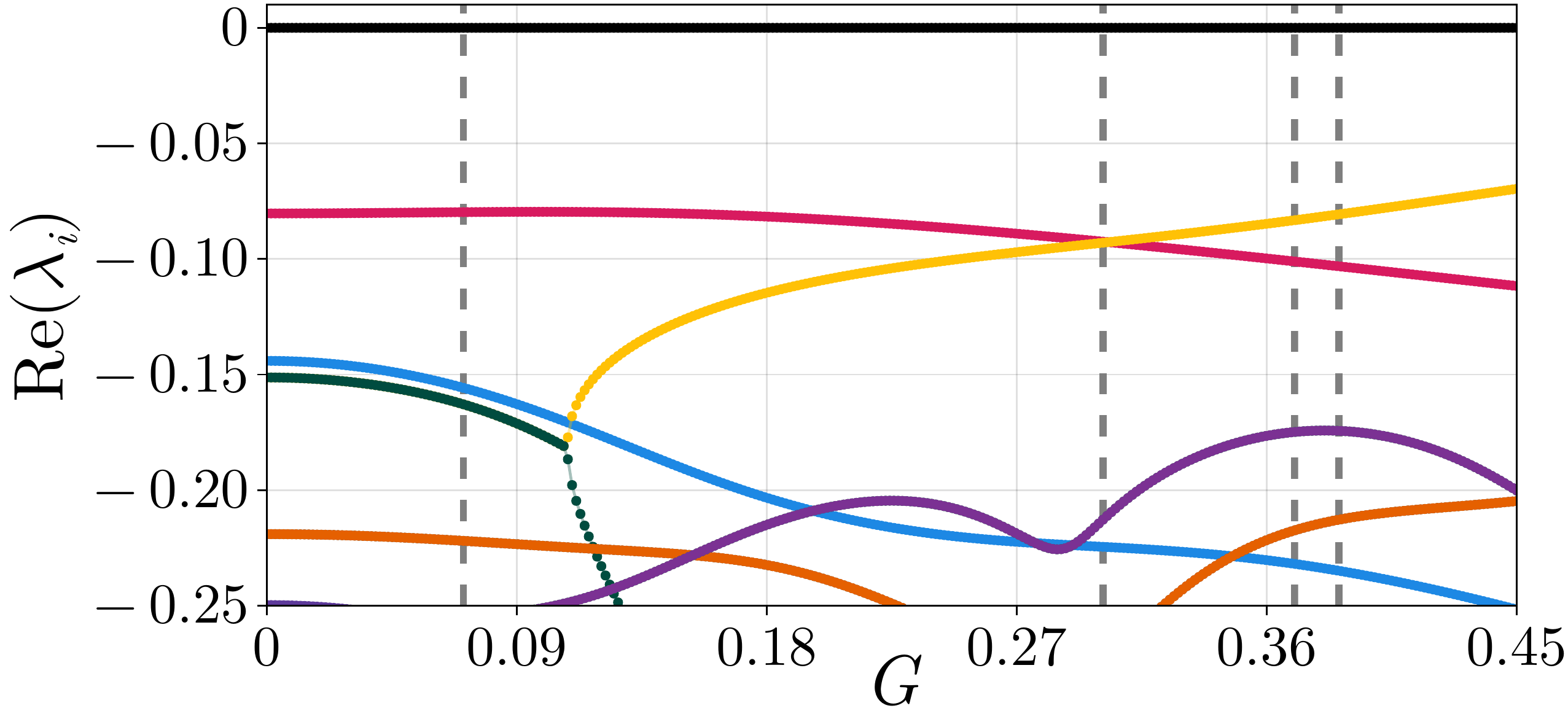}
  \caption{\textit{Liouvillian spectrum sensitivity to flow-topology transitions}. Real part of the low-lying Liouvillian eigenvalues $\text{Re}(\lambda_i)$ as a function of the drive amplitude $G$. The steady state corresponds to the eigenvalue $\lambda_0=0$ (black solid line). The vertical dashed lines denote the critical values associated with the local and global phase transitions identified in Fig.~\ref{fig:2}(f).}
  \label{fig:5}
\end{figure}

The molecule detects transitions that the Liouvillian spectrum misses. In particular, the real Liouvillian gap remains open [Fig.~\ref{fig:5}], even though, along the full sequence from $\bar{\alpha}_1^- S_1^-$ to $\alpha_2^-$, $\hat{\rho}_0$ evolves from an LC-like annulus to a bimodal Wigner distribution.
Local changes in the flow, such as $\bar{\alpha}_1^- S_1^-\mapsto\bar{\alpha}_2^- S_1^-$ leave the Liouvillian gap unchanged because they involve only unstable FPs; their signatures appear only at early times (large $|\mathrm{Re}(\lambda_i)|$), and are thus hard to extract from $\mathcal{L}$.
Likewise, the birth of repelling LCs between $\bar{\alpha}_2^- S_1^-\mapsto\alpha_2^- S_1^- \bar{S}_2^-$ produces only indirect spectral signatures, namely a crossing of the $\lambda_1$ and $\lambda_2$ decaying modes that inverts the slowest relaxation pathway in Fig.~\ref{fig:5} (see Appendix~\ref{sec:spectral_sign} for a calculation of the crossing point). 
Crucially, the molecule reveals that newly stabilized FPs are \textit{topologically protected} by the repelling LCs, acting as dynamical barriers in phase space. Only adjacent atoms can interact; in practice, sinks couple only to neighboring saddles, and LCs can be created or annihilated only when they are nested or adjacent to a saddle.
Because global reorganizations do not induce degeneracies in the Liouvillian (or Jacobian) spectrum, these transitions are not resolved by standard low-lying spectral diagnostics.

This conclusion should be understood with the usual caveat that, away from the strict semiclassical limit, quantum fluctuations deform the deterministic trajectories and induce leakage between the classical basins. The molecule is therefore not an exact invariant of the full quantum evolution. Rather, it remains a useful scaffold when the few-photon dynamics still retains enough memory of the underlying mean-field flow. We quantify this regime in Appendix~\ref{sec:validity_beyond}, where stochastic quantum trajectories are used to test how robustly the classical basin structure survives in the presence of quantum jumps and activation events.

\begin{figure*}[t]
    \centering
    \includegraphics[width=\textwidth]{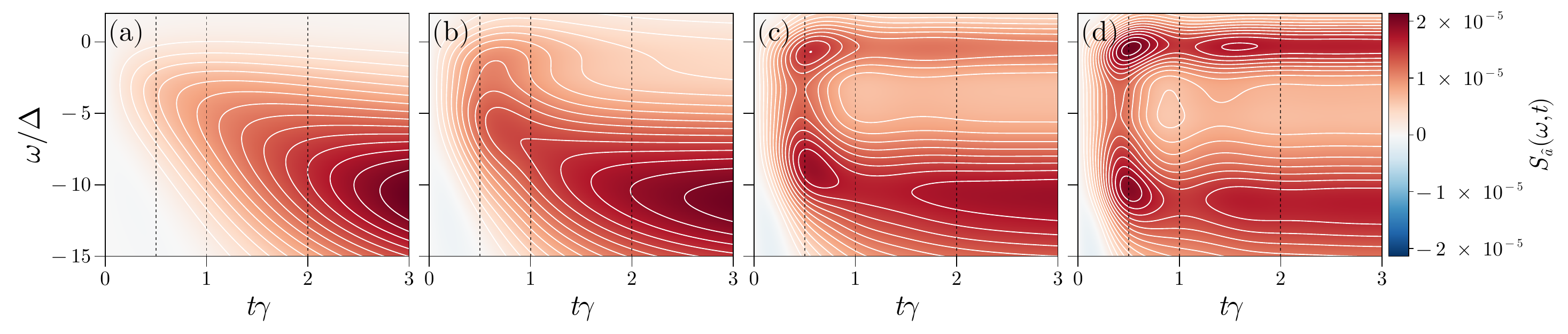}
    \caption{Time-resolved power spectrum for the parameter sets corresponding to
    Fig.~\ref{fig:3}(b)--(e). The two-photon drive amplitudes are $G=0, 0.2, 0.32, 0.38$, with $\Delta=0.05$, $\gamma = 0.1$, $U=-0.1$, and $\kappa_2=0.01$.}
    \label{fig:6}
\end{figure*}

\subsection{Experimental probes of flow-topological dynamics}
The physics described by Eq.~\eqref{Eq:liouvillian_parts} is accessible across several experimental platforms; here we highlight representative strengths. Nanomechanical systems enable reconstruction of the flow invariant~\cite{Dumont2024}, while trapped ions provide tunable nonlinearities and effective gain via laser-driven sidebands~\cite{liu_2025}. Levitated nanoparticles and polaritonic microcavities realize parametric instabilities and self-sustained LC dynamics through strong nonlinearities and resonant scattering~\cite{reisenbauer_non-hermitian_2024,liska_pt-like_2024,CarlonZambon2020}. Similar self-oscillatory regimes occur in cold-atom cavities~\cite{bohnet2012steady,Dreon2022}, superconducting circuits~\cite{astafiev2007single,Du2025,adinolfi2025}, and could be realized in quantum-acoustic systems~\cite{chu_quantum_2017}. The Wigner function is typically measured, for instance in superconducting cavity platforms, by dispersively coupling the oscillator to an ancilla qubit, applying controlled displacements, and reconstructing photon parity through Ramsey-based readout~\cite{Bild2023,Sun2014}.

While the Wigner function provides a direct visualization of the discussed phase-space structures, its reconstruction is experimentally demanding. A complementary route is provided by its quadrature marginals, which are directly related to experimentally measured probability distributions, see Appendix~\ref{app:wigner_marginals}. Beyond full tomography, one can also look for signatures of flow-topology transitions in simpler observables. A natural example is the time-resolved power spectrum
\begin{equation}
    S_{\hat a}(\omega,t)=\int_0^\infty d\tau\, e^{i\omega \tau}\,\langle \hat a^\dagger(t+\tau)\hat a(t)\rangle,
\end{equation}
where $\tau$ is the delay between the two field operators. This describes how the emitted field is distributed in frequency at a given evolution time $t$. In contrast to the usual steady-state spectrum, recovered for $t\to\infty$ when $G^{(1)}(t+\tau,t)=\langle \hat a^\dagger(t+\tau)\hat a(t)\rangle \to \langle \hat a^\dagger(\tau)\hat a(0)\rangle_{0}$ depends only on the delay, $S_{\hat a}(\omega,t)$ captures transient spectral features before the asymptotic regime is reached.

This observable is particularly useful in our setting because the transient dynamics is shaped by the phase-space structures encoded in the molecule. Rotations around attracting LCs, temporary trapping near unstable FPs, and exclusion from regions bounded by repelling cycles all produce characteristic time scales and oscillation frequencies in the correlator $G^{(1)}(t+\tau,t)$. These appear as transient peaks, broadened features, or redistributions of spectral weight in $S_{\hat a}(\omega,t)$.

We illustrate this observable in Fig.~\ref{fig:6}, for the same parameter sets used in the Wigner snapshots of Fig.~\ref{fig:3}. For the case corresponding to Fig.~\ref{fig:3}(b), where the state expands isotropically from the vacuum toward a stable LC, the spectrum shows the progressive emergence of a peak at the LC frequency, which becomes the dominant long-time feature. For the cases corresponding to Fig.~\ref{fig:3}(c)--(e), the transient evolution is more structured, reflecting motion near unstable FPs and LCs before the final relaxation to the steady state. In particular, for the phase corresponding to Fig.~\ref{fig:3}(c), one finds transient spectral weight associated with the temporary slowing down of the dynamics near unstable sources before the state is expelled toward the attracting LC. For the phases corresponding to Fig.~\ref{fig:3}(d) and Fig.~\ref{fig:3}(e), the growth of the LC peak is accompanied by additional low-frequency features that persist to longer times and are naturally associated with small oscillations around the stable FP attractors inherited from the Kerr-parametric-oscillator sector of the model.

At the same time, the distinction between the phases in Fig.~\ref{fig:3}(d) and Fig.~\ref{fig:3}(e), which are separated by the formation of a homoclinic connection, is not cleanly resolved by the time-resolved spectrum alone. The reason is that their main difference lies in the unstable phase-space structure and in the associated rearrangement of relaxation pathways, rather than in a sharply separated spectral scale. In this sense, the spectrum captures part of the flow-topology transition, but not all of it. The time-resolved spectrum should therefore be viewed not as a complete classifier of the molecule, but as an experimentally accessible probe of selected transient pathways. It captures the emergence of LC motion and additional slow features associated with temporary trapping near unstable structures, while the full topological distinction requires phase-space information.

\section{Conclusion and Outlook}

Our framework shows that flow-topology transitions reorganize the fluctuation-relaxation pathways of the quantum dynamics. We demonstrated this in a driven-dissipative Kerr parametric oscillator with nonlinear gain and loss, where changes in the molecule invariant reshape the transient motion of quantum states through phase space, even when they leave weak signatures, or none at all, in the low-lying Liouvillian spectrum. More broadly, by introducing a constructive procedure to build the invariant from fixed points, limit cycles, separatrices, and their chirality, the approach provides a practical route to identify which phase-space structures guide quantum relaxation and which topological transitions are allowed. This suggests that the same framework should also apply to rare activation paths between metastable states, enabling predictions for quantum activation pathway topology in non-equilibrium phases~\cite{marthaler_switching_2006,Macieszczak2016}. 
The construction assumes a phase space compactified into a sphere, where the molecule fits the Poincaré–Hopf theorem: the attractor, repeller, and saddle indices sum up to the Euler characteristic; see Appendix~\ref{sec:morse_smale}. Extending it to tori or flux-threaded Brillouin zones would alter these constraints toward linking flow topology with band topology in nonlinear topological photonics~\cite{smirnova_nonlinear_2020,szameit_discrete_2024}, potentially revealing additional invariants and phases of matter.
Sensitivity of relaxation pathways to global bifurcations further enables applications paralleling those of local bifurcations, such as bifurcation-based amplification and sensing~\cite{vijay_invited_2009}, engineered bosonic codes operating near critical points~\cite{Gravina_2023}, and quantum-control landscape engineering~\cite{beato_toward_2025}.

\section*{Acknowledgments}
The authors thank O. Zilberberg for bringing to our attention the work by Mutschler et al.~\cite{Mutschler2025}, which develops a framework for classifying limit cycles in nonlinear dynamical systems. The authors also thank O. Zilberberg and D. Porras for valuable discussions.

We acknowledge funding from the Ramón y Cajal program (RYC2023-043827-I), funded by MICIU/AEI (10.13039/501100011033) and FSE+, and from the ``María de Maeztu'' Programme for Units of Excellence in R\&D (CEX2023-001316-M), funded by the Spanish Ministry of Science, Innovation and Universities. J.d.P. also acknowledges support from the Proyectos de Generación de Conocimiento program, Grant No. PID2024-158923NA-I00, funded by MICIU/AEI (10.13039/501100011033) and FEDER, UE.

\section*{Data Availability}

The data supporting the findings of this article, together with the scripts used to generate the figures, are openly available in Zenodo~\cite{GomezZenodo2026}.

\appendix


\section{Local transitions: Semiclassical fixed-point analysis}\label{sec:local_trans}

\subsection{Mean field flow}

The dynamics of the system are governed by the Lindblad master equation for the density matrix $\hat{\rho}$:
\begin{equation}
    \partial_t \hat{\rho} = -i [\hat{\mathcal{H}},\hat{\rho}] + \kappa_+ \mathcal{D}[\hat{a}^\dagger]\hat{\rho} + \kappa_- \mathcal{D}[\hat{a}]\hat{\rho} + \kappa_2 \mathcal{D}[\hat{a}^2]\hat{\rho},
\end{equation}
where the Hamiltonian in the rotating frame is given by 
\begin{equation}
\hat{\mathcal{H}} = -\Delta \hat{a}^\dagger\hat{a}+ \frac{U}{2} \hat{a}^{\dagger 2}\hat{a}^2 + \frac{G}{2} (\hat{a}^{\dagger 2}+ \hat{a}^2).
\end{equation}
In the limit of large photon numbers, we apply the mean-field approximation, e.g., $\langle \hat{a}^\dagger \hat{a} \hat{a} \rangle \approx |\alpha|^2 \alpha$, where $\alpha \equiv \langle \hat{a} \rangle$. The equation of motion for the complex amplitude $\alpha$ reads:
\begin{equation}
    i \dot{\alpha} = \left[-\Delta + i \frac{\gamma}{2} + (U - i \kappa_2) |\alpha|^2\right] \alpha + G \alpha^*,
\end{equation}
with the effective linear rate $\gamma \equiv \kappa_+ - \kappa_-$. Decomposing the amplitude into real quadratures $\alpha = u + i v$ and defining the intensity $r^2 = u^2 + v^2$, we obtain the planar flow presented in Eq.~\eqref{Eq:SemiclassicalEOM}:
\begin{equation}
\label{eq:flow_cartesian}
\begin{aligned}
    \dot{u} &= \left( \frac{\gamma}{2}-\kappa_2 r^2\right)u - (\Delta+G-Ur^2)v, \\
    \dot{v} &= \left(\frac{\gamma}{2}-\kappa_2r^2 \right)v +(\Delta-G-Ur^2)u.
\end{aligned}
\end{equation}

\subsection{Fixed points and phase boundaries}
It is instructive to express the dynamics in polar coordinates $\alpha = r e^{i\phi}$, which yields:
\begin{subequations}
\begin{align}
    \dot{r} &= r \left(\frac{\gamma}{2}- \kappa_2 r^2-G \sin(2\phi)\right), \\
    \dot{\phi} &= \Delta - Ur^2 - G\cos(2\phi).
\end{align}
\end{subequations}
Note that the angular equation is valid for $r \neq 0$, while the origin $r=0$ is a singular point of the polar map but a well-defined FP of the Cartesian flow.

The FPs $(r_\star, \phi_\star)$ obey $\dot{r}=\dot{\phi}=0$. The origin $r_\star=0$ (or $\alpha=0$) is a trivial solution for all parameter values. Due to the $\mathbb{Z}_2$ symmetry of the Liouvillian under the parity transformation $\hat{a} \to -\hat{a}$, nontrivial FPs appear in pairs $(r_\star, \phi_\star)$ and $(r_\star, \phi_\star + \pi)$, corresponding to $\pm \alpha_\star$. Consequently, the total number of FPs is always odd. For nontrivial solutions ($r_\star > 0$), the stationarity conditions imply:
\begin{equation}
    G \sin (2\phi_\star) = \frac{\gamma}{2}-\kappa_2 r_\star^2, \quad G \cos(2\phi_\star) = \Delta - Ur_\star^2.
\end{equation}
Squaring and summing these equations eliminates the phase $\phi_\star$, leading to a self-consistency condition for the intensity $I_\star \equiv r_\star^2$:
\begin{equation}
    G^2 = \left(\frac{\gamma}{2}-\kappa_2 I_\star\right)^2+(\Delta-U I_\star)^2.
\end{equation}
This can be rearranged into a quadratic equation for the intensity, $S I_\star^2 - A I_\star + C = 0$, with coefficients defined as~\cite{rand2012lecture}:
\begin{equation}
\label{eq:S_A_B_def}
    S \equiv \kappa_2^2 + U^2, \quad 
    A \equiv 2\Delta U + \gamma \kappa_2, \quad 
    C \equiv \frac{\gamma^2}{4} + \Delta^2 - G^2.
\end{equation}
The solutions are formally given by:
\begin{equation}\label{eq:FP_intensities}
    I_{\star,\pm} = \frac{A \pm \sqrt{\mathcal{D}}}{2S}, \quad \text{with } \mathcal{D} = 4S G^2 - B^2,
\end{equation}
where we have introduced the auxiliary parameter $B \equiv U\gamma - 2\kappa_2\Delta$ and a discriminant $\mathcal{D} = A^2 - 4SC$~\cite{minganti_spectral_2018,Dutta2025}.

Physical FPs correspond to real, positive intensities $I_\star > 0$. We classify the solutions based on the drive strength $G$ and the sign of $A$ in Eq.~\eqref{eq:S_A_B_def}. We define two critical drive amplitudes, namely~\cite{haga_oscillating-mode_2024}
\begin{equation}\label{eq:G_bif}
    G_{\min} \equiv \frac{|B|}{2\sqrt{S}}, \qquad G_{\star} \equiv \frac{\sqrt{A^2+B^2}}{2\sqrt{S}}.
\end{equation}
Here, $G_{\min}$ corresponds to the threshold where the discriminant becomes non-negative ($\mathcal{D} \ge 0$), and $G_\star$ corresponds to the point where one solution branch crosses zero ($C=0$). The FP landscape is divided into three distinct regions:

\begin{enumerate}
    \item \textbf{Single Trivial Attractor (1 FP):}
   For $|G|<G_{\min}$, no real nontrivial solution exists; only the trivial fixed point remains.
    
    \item \textbf{Five-Solution Region (5 FPs):}
    This phase requires two distinct positive roots $I_{\star,\pm}>0$, which occurs when $A>0$ and $C>0$ (i.e., $|G|<G_\star$). For $A>0$ and $G_{\min}<|G|<G_\star$, the system supports five FPs: the origin and two nontrivial pairs.
    
    \item \textbf{Three-Solution Region (3 FPs):}
    This phase occurs when exactly one root $I_\star>0$. For $A>0$, this requires a negative product of roots ($|G|\ge G_\star$). For $A<0$, a real discriminant and a positive root likewise require a negative product, again giving $|G|\ge G_\star$. Thus, for any $A$, the condition $|G|\ge G_\star$ yields exactly three FPs: the origin and one nontrivial pair.
\end{enumerate}

These analytic bounds $G_{\min}$ and $G_\star$ in Eq.~\eqref{eq:G_bif} thus define the transition lines (bifurcations) observed in the phase diagrams of Fig.~\ref{fig:2}. Specifically, $G_{\min}$ marks the saddle-node bifurcation creating the 5-FP region, while $G_\star$ marks the supercritical pitchfork bifurcation where the origin loses stability.

\subsection{Linear stability analysis and Jacobian}
The stability is governed by the Jacobian matrix of the flow in Eq.~\eqref{eq:flow_cartesian}, $J(\mathbf{r}) \equiv \partial(f_u, f_v)/\partial(u,v)$.
Differentiating the flow equations, the Jacobian evaluated at an arbitrary point is
\begin{widetext}
\begin{equation}
J(u,v) = \begin{pmatrix}
\frac{\gamma}{2} - \kappa_2(r^2+2u^2) + 2Uuv & -(\Delta+G) + Ur^2 + 2Uv^2 - 2\kappa_2 uv \\
\Delta - G - Ur^2 - 2Uu^2 - 2\kappa_2 uv & \frac{\gamma}{2} - \kappa_2(r^2+2v^2) - 2Uuv
\end{pmatrix}.
\end{equation}
\end{widetext}
The stability analysis follows the standard trace-determinant plane~\cite{rand2012lecture}, where the trace $\mathfrak{T} \equiv \operatorname{tr}(J)$ and determinant $\mathfrak{D} \equiv \det(J)$, read
\begin{align}\label{eq:trace_determinant}
    \mathfrak{T}(r^2) &= \gamma - 4\kappa_2 r^2, \\
    \mathfrak{D}(r^2) &= r^2(4Sr^2 - 2A),
\end{align}
with $S$ and $A$ defined in Eq.~\eqref{eq:S_A_B_def}. The character of the FP depends on the Jacobian discriminant $\delta_J \equiv \mathfrak{T}^2 - 4\mathfrak{D}$:
\begin{equation}
    \delta_J(r^2) = \gamma^2 + 16Ur^2(\Delta - Ur^2).
\end{equation}

Substituting the explicit solutions \eqref{eq:FP_intensities} into Eq.~\eqref{eq:trace_determinant} simplifies to
\begin{equation}
    \mathfrak{D}(I_{\star,\pm}) = 2 I_{\star,\pm} \left( 2S I_{\star,\pm} - A \right) = \pm 2 I_{\star,\pm} \sqrt{\mathcal{D}}.
\end{equation}
Since physical FPs require $I_{\star} > 0$ and the existence discriminant $\mathcal{D} \ge 0$, the sign of the determinant is fixed solely by the solution branch:
\begin{equation}
    \mathfrak{D}(I_{\star,+}) > 0, \qquad \mathfrak{D}(I_{\star,-}) < 0.
\end{equation}
The ``$-$'' branch (when it exists physically) contains only saddle points. The ``$+$'' branch is always an antisaddle (sink or source), and its stability is determined solely by the trace $\mathfrak{T}$.

\textit{Hopf Bifurcation:} A Hopf instability occurs when the trace vanishes, $\mathfrak{T}(I_{\star,+}) = 0$. This requires $4\kappa_2 I_{\star,+} = \gamma$. Solving for $G$ yields the critical Hopf drive:
\begin{equation}
\label{eq:G_Hopf}
G_{\mathrm{H}} = \frac{1}{2\sqrt{S}}\sqrt{B^{2}+\left(\frac{S\gamma}{2\kappa_{2}}-A\right)^{2}}.
\end{equation}
At $|G| = G_{\mathrm{H}}$, the system undergoes a subcritical Hopf bifurcation, giving rise to an unstable LC.

\textit{Node-Focus Transition:} The transition from monotonic (node) to oscillatory (focus) approach to the FP occurs when $\delta_J(I_{\star,+}) = 0$. Solving this condition leads to the threshold:
\begin{equation}
\label{eq:FocusNodeTransition}
G_{\mathrm{F}}^{(\pm)} = \frac{1}{2\sqrt{S}}
\sqrt{B^{2}+\left[\frac{S}{2U}\left(2\Delta\pm\sqrt{\gamma^{2}+4\Delta^{2}}\right)-A\right]^{2}},
\end{equation}
which also manifests a level crossing of Liouvillian eigenmodes in Fig.~\ref{fig:5}. For $U=0, \kappa_2 \neq 0$ (2DPO), $\delta_J = \gamma^2 > 0$, implying the FP is always a node.

\paragraph{Stability of the Origin:} Linearizing around the trivial solution $\alpha=0$ ($r=0$), the eigenvalues are 
\begin{equation}
    \varepsilon_{\pm} = \frac{\gamma}{2} \pm \sqrt{G^2 - \Delta^2}.
\end{equation}
We distinguish two regimes split by a second order exceptional point (EP) at $G=\Delta$:

\begin{itemize}[leftmargin=10pt, nosep]
\item $|G| < |\Delta|$: The eigenvalues are complex conjugates, $\varepsilon_{\pm} = \gamma/2 \pm i\Omega$. The origin is a stable focus if $\gamma < 0$ (loss dominates gain) and unstable source if $\gamma > 0$.

\item $|G| > |\Delta|$: The eigenvalues become real. If  $|\gamma| > 2\sqrt{G^2 - \Delta^2}$, the origin remains a node (stable if $\gamma < 0$). Else, if $|\gamma| < 2\sqrt{G^2 - \Delta^2}$, one eigenvalue becomes positive, turning the origin into a saddle point (pitchfork bifurcation).
\end{itemize}

\section{Structural stability and topological transition rules}\label{sec:morse_smale}
\subsection{Morse--Smale conditions and structural stability}
Structural stability expresses that a continuous dynamical system, governed by a flow, varies smoothly under small perturbations, remaining topologically equivalent to the original one. A structurally stable system retains its qualitative behavior under small perturbations of its parameters: FPs do not change type, LCs persist (though their positions and periods may shift), and separatrices deform smoothly without altering their connectivity. In two-dimensional systems, this robustness is ensured when the flow satisfies the Morse–Smale conditions. These require that (i) all FPs and periodic orbits are hyperbolic, that (ii) only finitely many of them exist, and that (iii) their stable and unstable manifolds intersect transversely~\cite{PEIXOTO1962101, villa2025topological}. Hyperbolicity ensures the dynamics are decisive: at any equilibrium or closed loop, the system strictly attracts or repels nearby trajectories, with no neutral directions where behavior is ambiguous.

When only FPs are present, these conditions guarantee that the flow decomposes cleanly into basins separated by saddle manifolds, with no recurrent behavior beyond the FPs themselves. With LCs this picture generalizes naturally: hyperbolic cycles act as periodic attractors or repellers. The transversality condition ensures that the stable and unstable manifolds of saddles and cycles do not form tangencies, thereby forbidding chaotic dynamics and guaranteeing that the phase space remains partitioned into well-defined basins of attraction. In our model, away from the bifurcation lines in the phase diagrams, the semiclassical flow satisfies these Morse–Smale requirements, ensuring structural stability throughout each phase and validating the use of a topological invariant to classify them~\cite{soriente_distinctive_2021}.

\subsection{Local and global bifurcation rules of the molecule invariant}
The global topology of the flow is constrained by the Poincaré–Hopf index theorem on the sphere, which requires the sum of the indices of all singularities (FPs) to satisfy $\sum_i \text{ind}(x_i) = 2$~\cite{nakahara2018geometry}. In terms of our atoms: $N_\circ + N_\bullet - N_\times = 2,$
where $N_\circ$, $N_\bullet$, and $N_\times$ are the numbers of sources, sinks, and saddles, respectively. Note that hyperbolic LCs ($S, \bar{S}$) carry a Poincaré index of 0, so they can be created or destroyed without altering this sum, although their appearance fundamentally changes the connectivity of the molecule. Local events such as saddle–node bifurcations, supercritical and subcritical Hopf bifurcations, and chirality flips preserve the total index while changing the molecule in controlled ways consistent with these adjacency constraints.

The molecule invariant imposes topological rules on the allowed phase transitions. Because physical bifurcations correspond to specific operations on the graph (such as edge contraction or vertex merging), only atoms that are adjacent in the molecule can interact. This means that a FP can only undergo a bifurcation with a saddle if they share a direct bond (i.e. separatrix connection), and two LCs can collide and annihilate only if they are nested, which makes them adjacent in the invariant. This structure also provides a mechanism of topological protection: an attractor is dynamically separated by other atoms if they are not adjacent (e.g., a repelling LC acting as a barrier as in Fig.~\ref{fig:3}), ensuring that local and global reorganizations must propagate through the graph neighbor by neighbor rather than jumping between disconnected atoms.

\begin{figure*}[t]
  \centering
  \includegraphics[width=\linewidth]{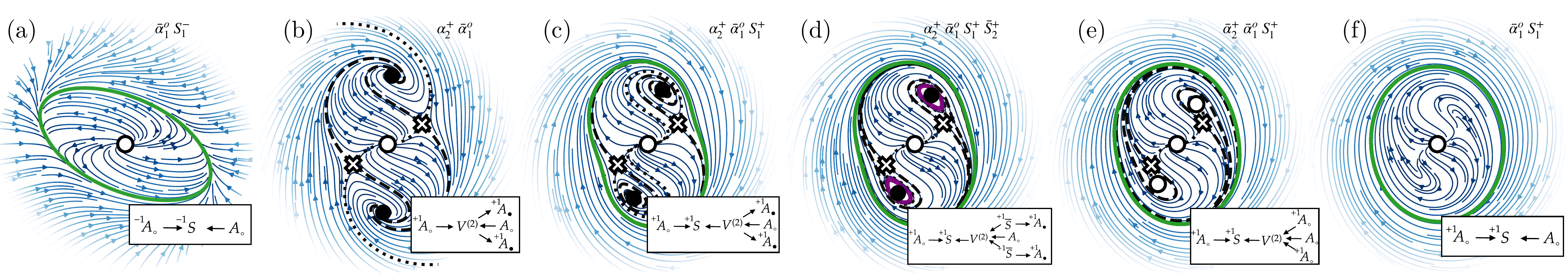}
  \caption{\textit{Semiclassical flows for phases with five FPs}. (a)--(f) Mean-field vector flows in the $(u,v)$ plane corresponding to the sequence of topological transitions driven by the Kerr nonlinearity $U/\gamma \in \left\{0.01, 0.2, 0.28, 0.32, 0.4, 0.6 \right\}$ at fixed two-photon drive amplitude $G/\gamma=0.4$ (other parameters as in Fig.~\ref{fig:2}). Green (purple) closed curves denote attracting (repelling) LCs $S$ $(\bar{S})$, black (white) dots represent point attractors (repellers) and dashed (dotted) lines indicate the stable (unstable) manifolds of the saddle points (crosses).}
  \label{fig:S1}
\end{figure*}

\section{Global reconnections in the 5FP region}\label{sec:global_5FP}
We now focus on the five-fixed-point region of the phase diagram, where the coexistence of several fixed points and limit cycles produces a particularly rich sequence of flow-topology transitions. We analyze this sequence by varying the ratio $U/\gamma$, which reveals a rich interplay between local instabilities and global reorganizations of the phase space. Starting from a self-oscillatory phase characterized by a single unstable source surrounded by an attracting LC, the system first undergoes a double saddle-node-on-invariant-circle (SNIC) bifurcation $\bar{\alpha}_1^- S_1^- \mapsto \alpha_2^+ \bar{\alpha}_1^o$ [Figs.~\ref{fig:S1}(a) and \ref{fig:S1}(b)]. In this global event, two pairs of saddle-node FPs nucleate directly upon the invariant circle of the LC. Unlike a standard local bifurcation, the SNIC destroys the periodic orbit not by reducing its amplitude to zero, but by diverging its period as the flow becomes arrested at the emerging stable nodes \cite{glendinning1994stability}.

As the parameter is increased further, the global manifolds of the newly created saddles undergo a heteroclinic reconnection $\alpha_2^+ \bar{\alpha}_1^o \mapsto \alpha_2^+ \bar{\alpha}_1^o S_1^+$ [Figs.~\ref{fig:S1}(b) and \ref{fig:S1}(c)]. The unstable manifolds of the saddles, which initially terminated at the local sinks, reconnect with the stable manifolds of their symmetric counterparts. This global rewiring births a large, enclosing attracting LC that surrounds the entire fixed-point complex, marking the onset of a phase where local stationarity coexists with large-amplitude free-phase oscillatory dynamics.

The internal structure of this coexistence phase subsequently evolves through homoclinic bifurcations $\alpha_2^+ \bar{\alpha}_1^o S_1^+ \mapsto \alpha_2^+ \bar{\alpha}_1^o S_1^+ \bar{S}_2^-$ [Figs.~\ref{fig:S1}(c) and \ref{fig:S1}(d)]. Here, the stable separatrix of each saddle contracts and collides with the saddle itself. This collision generates a repelling LC surrounding each local sink, effectively creating a dynamical barrier that isolates the basin of attraction of the FPs from the outer flow. These repelling cycles, as the nonlinearity evolves, shrink and eventually collapse onto the stable nodes in a subcritical Hopf bifurcation $\alpha_2^+ \bar{\alpha}_1^o S_1^+ \bar{S}_2^- \mapsto \bar{\alpha}_2^+ \bar{\alpha}_1^o S_1^+$ [Figs.~\ref{fig:S1}(d) and \ref{fig:S1}(e)]. This event transfers the instability to the FPs, converting the sinks into unstable sources and destroying the repelling cycles.

Finally, the remaining complex of unstable FPs—now consisting of sources and saddles—is annihilated via an inverse saddle-node bifurcation $\bar{\alpha}_2^+ \bar{\alpha}_1^o S_1^+ \mapsto \bar{\alpha}_1^o S_1^+$ [Figs.~\ref{fig:S1}(e) and \ref{fig:S1}(f)]. The disappearance of these critical points removes the obstruction to the central flow, allowing trajectories to spiral outward once more toward the large attracting LC.

\section{Spectral signatures of local transitions}\label{sec:spectral_sign}
\subsection{Liouvillian and oscillating-mode gaps}

Each eigenvalue $\lambda_j$ of the Liouvillian $\tilde{\mathcal{L}}$ (Eq.~\eqref{eq:general_eq}) has a real part $\mathrm{Re}(\lambda_j)<0$ that sets the decay rate of a transient mode and an imaginary part $\mathrm{Im}(\lambda_j)$ that determines its oscillation frequency. Stationary states correspond to modes with $\mathrm{Re}(\lambda_j)=0$, and in the case of LCs they may carry nonzero $\mathrm{Im}(\lambda_j)$, reflecting persistent oscillations and the spontaneous breaking of continuous time-translation symmetry.~\cite{minganti_spectral_2018,Dutta2025}

A closure of the \textit{Liouvillian} gap, defined as the smallest nonzero decay rate, i.e.,
\begin{equation}
    \Delta_{\mathrm L}=\min_{i\neq 0}\big[-\mathrm{Re}(\lambda_i)\big],
\end{equation}
signals critical slowing down and a sharp transition in $\hat{\rho}_0$: relaxation toward the steady state becomes arbitrarily slow as the dominant decay mode vanishes. However, dynamical phase transitions can also occur in the transient regime without closing $\Delta_\mathrm{L}$. This is captured by the so-called oscillating-mode gap~\cite{haga_oscillating-mode_2024}, defined from the least-damped oscillatory modes:
\begin{equation}
    \Delta_{\mathrm{OM}}=\min_{i:\,\mathrm{Im}(\lambda_i)\neq 0}\big[-\mathrm{Re}(\lambda_i)\big].
\end{equation}
This quantity is sensitive to EPs in the Liouvillian excitation spectrum. At a chirality-loss transition, such as the 2DPO $\to$ KPO crossover, two complex eigenvalues cross on the real axis. Physically, this marks a sharp transition from underdamped, spiraling relaxation to overdamped decay. Across this transition $\Delta_{\mathrm{OM}}$ varies, while $\Delta_\mathrm{L}$ remains open. Distinguishing Hamiltonian-level EPs from Liouvillian EPs induced by quantum jumps, the EPs relevant here are inherited from the semiclassical fluctuation dynamics around the FP rather than from the jump sector of the Liouvillian \cite{minganti_quantum_2019}.

Global flow-topology changes, instead, reroute relaxation trajectories (e.g. due to the annihilation of nested LCs of opposite stability) without modifying the low-lying Liouvillian spectrum. Both $\Delta_{\mathrm L}$ and $\Delta_{\mathrm{OM}}$ remain essentially unchanged, while the transient Wigner dynamics reorganize abruptly. Figure~\ref{fig:5} illustrates this mismatch, showing clear dynamical transitions in the time-dependent Wigner function that are invisible to the Liouvillian gaps.

\subsection{Exceptional point and chirality loss in fixed-point phases}

\begin{figure}[!t]
     \centering
    \includegraphics[width=\linewidth]{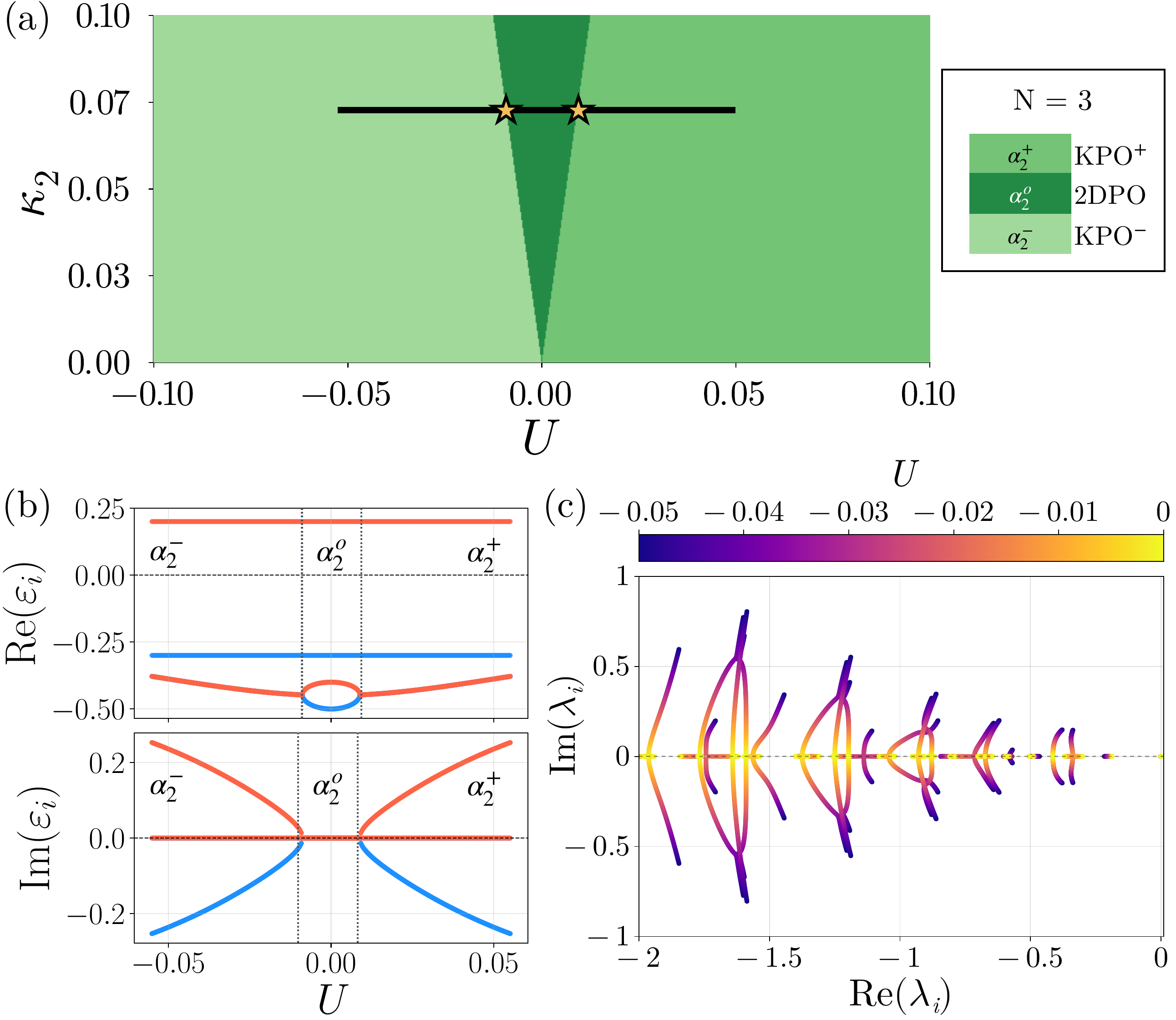}
    \caption{\textit{Chirality loss transition.} (a) Phase diagram in the $(U,\kappa_2)$ plane showing the transition between the chiral (underdamped) KPO $(\alpha^{\pm}_2)$ and the achiral (overdamped) 2DPO $(\alpha^o)$ phases. Stars mark the exceptional points along the $\kappa_2=0.07$ cut line. (b) Real and imaginary parts of the Jacobian eigenvalues $\varepsilon_i$ along the cut, showing the coalescence of eigenvalues in the overdamped region. (c) Evolution of the Liouvillian spectrum $\lambda_i$ across the transition (from $U=-0.05$ to $0$), where oscillating modes (darker colors) collapse onto the real axis (lighter colors).}
    \label{fig:S2}
\end{figure}

Figure~\ref{fig:1}(b) highlights a key point: two stationary phases can look essentially identical at the level of the steady state, yet differ in their underlying dynamics. In both cases, the semiclassical flow features two attractors separated by a saddle, producing a bimodal Wigner function. However, the corresponding phase-space flows [Fig.~\ref{fig:1}(d)] are topologically inequivalent, with signatures in the Jacobian fluctuation spectrum and in the mapping between Liouvillian spectra shown in Fig.~\ref{fig:1}(c). Namely, in the FP phases associated with the KPO/2DPO crossover [cf. the phase diagram in Fig.~\ref{fig:S2}(a)], the chirality-loss transition $3\alpha_2^- \mapsto 3\alpha_2^o$ corresponds to an exceptional point of the Jacobian governing fluctuations around the attractor. An explicit analysis of the Jacobian eigenvalues as functions of $\kappa_2/U$ and $G/\gamma$, at zero detuning $\Delta=0$, leads to the EP condition in Eq.~\eqref{eq:FocusNodeTransition}, namely
\begin{equation}
    \frac{\kappa_2}{U} = 2 \pm \sqrt{16\left(\frac{G}{\gamma}\right)^2 - 1},
\end{equation}
where the Jacobian eigenspectrum coalesces, as noted in the eigenvalues in Fig.~\ref{fig:S2}(b) and corresponding eigenvectors (not shown). At this line, the dynamics undergoes a transition from underdamped (finite chirality / symplectic norm) to overdamped (zero chirality) behavior. This change is reflected in the oscillating-mode gap of the Liouvillian but does not close the steady-state gap; see Fig.~\ref{fig:S2}(c).

\subsection{Hopf bifurcation and onset of the RVdP LC}

In the RVdP limit, linearizing the Liouvillian around the trivial steady state reveals a Hopf instability: a conjugate pair of eigenvalues crosses the imaginary axis, which can be obtained from the Jacobian analysis \cite{Dutta2025}. At this point, the real part of the leading decay mode vanishes, the Liouvillian gap closes, and the system develops a finite oscillation frequency set by the imaginary part of this pair, in agreement with the RVdP behavior shown in Fig.~\ref{fig:1}.

\begin{figure}[!t]
     \centering
    \includegraphics[width=\linewidth]{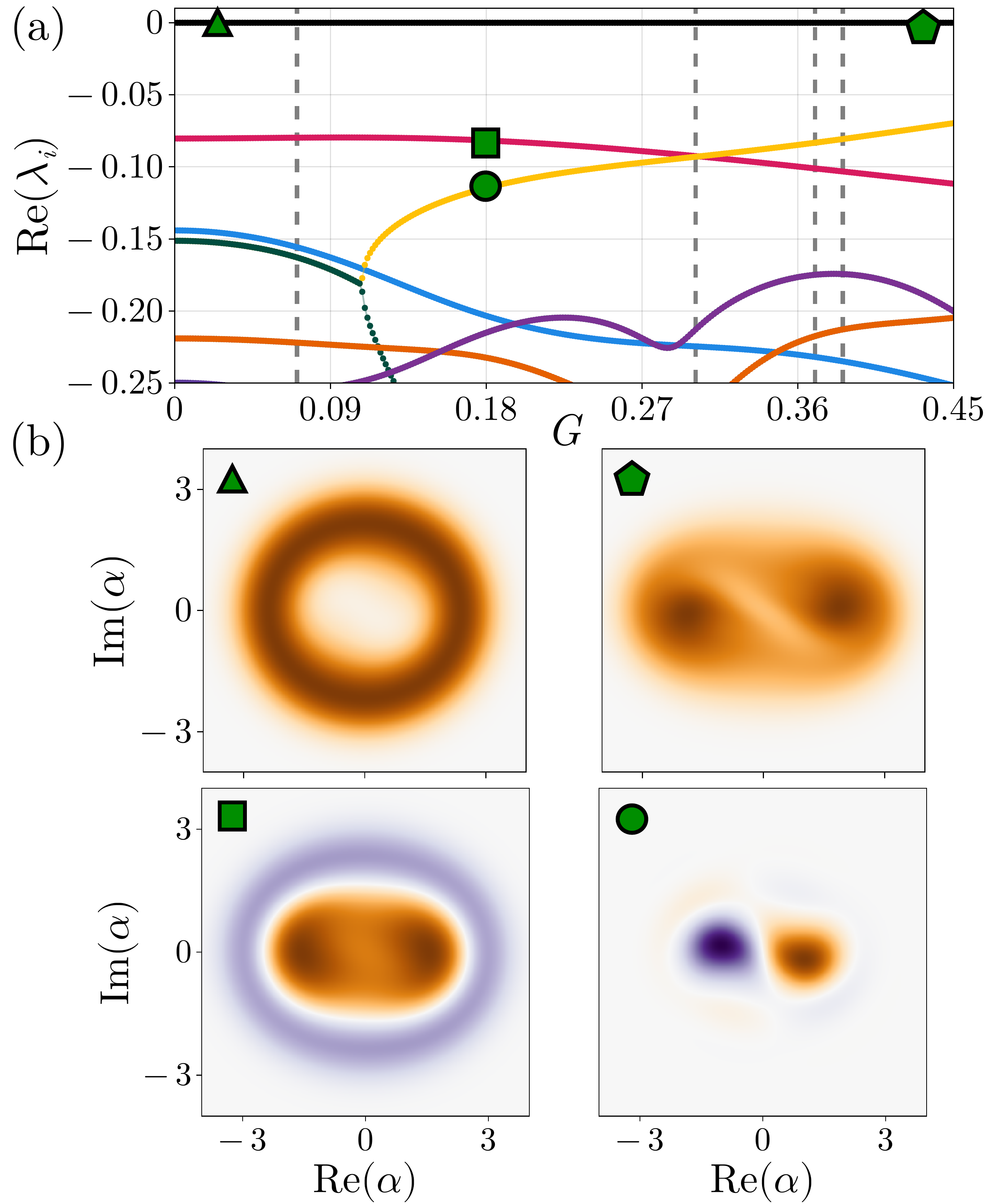}
    \caption{\textit{Liouvillian spectrum and low-lying decay eigenmodes.} (a) Real part of the Liouvillian eigenvalues as in Fig.~\ref{fig:5}. (b) Wigner function of the eigenmodes 0, 1, and 2 marked with symbols in panel (a).}
    \label{fig:S3}
\end{figure}

For the sequence of phases with LCs in Fig.~\ref{fig:3}(a), the critical drive $G_{\mathrm H}$ given by Eq.~\eqref{eq:G_Hopf} instead marks an inversion of the slowest relaxation pathway. It is obtained by evaluating the Jacobian at the relevant fixed point and relating its eigenvalues to the low-lying Liouvillian modes $\lambda_1$ and $\lambda_2$. Requiring their decay rates to coincide yields the analytic expression quoted above. Figures~\ref{fig:S3}(a) and \ref{fig:S3}(b) compare this prediction with numerical spectra, showing the level crossing of $\mathrm{Re}(\lambda_{1,2})$ at $G_{\mathrm H}$.

\section{Validity beyond the semiclassical limit}\label{sec:validity_beyond}

\begin{figure}[!t]
    \centering
    \includegraphics[width=\linewidth]{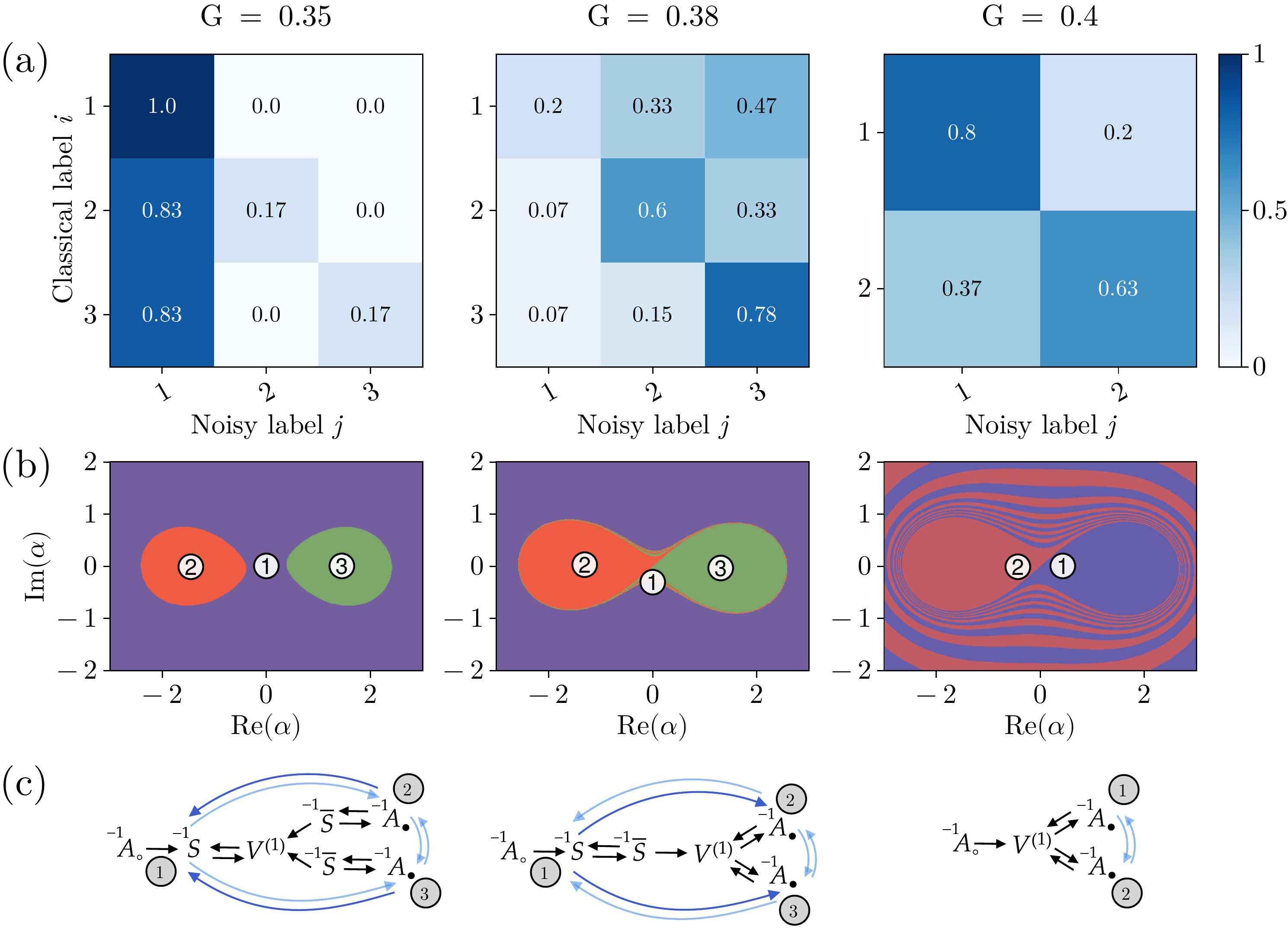}
    \caption{\textit{Robustness analysis of the molecule invariant}. (a) Confusion matrices $C_{ji}$ for three representative multistable phases at $G=0.35, 0.38, 0.40$. The corresponding values of $\epsilon$, namely $0.553$, $0.473$, and $0.285$, respectively, reflect the increasing fidelity with which the flow topology guides the quantum dynamics. (b) Corresponding classical basins of attraction in phase space. (c) Coarse-grained molecule invariants. Black arrows denote the molecule invariant, while blue arrows indicate fluctuation-induced interbasin transfer inferred from the confusion matrices in panel (a). Darker (lighter) blue corresponds to larger (smaller) transfer probability.}
    \label{fig:S4}
\end{figure}

As one moves deeper into the quantum regime, fluctuations progressively blur the classical separatrices and activate switching between basins of attraction. This behavior is naturally captured by the evolution equation of the Wigner function,
\begin{align}
\label{eq:Wigner_Ev}
\partial_t W(\boldsymbol r,t)
={}&
-\nabla \!\cdot\! \bigl[\mathbf f(\boldsymbol r)\, W(\boldsymbol r,t)\bigr]
\notag\\
&+
\sum_{i,j}
\partial_i \partial_j
\bigl[D_{ij}(\boldsymbol r)\, W(\boldsymbol r,t)\bigr]
\notag\\
&+
\mathcal Q[W] .
\end{align}
where $\boldsymbol r=(u,v)^T\), \(D_{ij}(\boldsymbol r)\) denotes the phase-space diffusion matrix generated by the quantum noise in phase space, and $\mathcal Q[W]$ collects higher-order quantum corrections beyond the drift-diffusion terms. The first term describes the deterministic flow generated by the semiclassical field $\mathbf f(\boldsymbol r)$, the second accounts for the spreading induced by quantum noise, and $\mathcal Q[W]$ collects higher-order quantum corrections~\cite{groenewold1946principles, Moyal1949, zachos2005quantum, Venkatraman2022}. The evolution of $W$ naturally raises the question of how far the molecule invariant continues to provide a useful organizing scaffold once the typical occupation is reduced to only a few quanta.

The transient Wigner dynamics in Fig.~\ref{fig:3} show that, even in the intermediate nonlinear regime $U/\kappa\sim 1$, the evolution still follows the broad structure of the underlying semiclassical flow: probability tends to move along the classical trajectories, accumulate near stationary sets, and reorganize when a global bifurcation reshapes the basin structure. At the same time, quantum fluctuations blur the classical boundaries and enable leakage between neighboring basins through both quantum activation and tunneling.

A useful way to connect the classical deterministic flow to the quantum regime is through quantum trajectories, namely individual stochastic realizations of the open-system dynamics. Classically, the flow $\mathbf f(\boldsymbol r)$ assigns a unique trajectory, and hence a unique basin of attraction, to each initial condition. In the quantum problem, each realization still follows an individual path shaped by the same underlying flow, but now with additional fluctuations and jumps that can progressively blur the classical basin structure. This makes quantum trajectories a natural tool to quantify how much of the molecule remains beyond the semiclassical limit. In the quantum-jump description, the conditioned state $\ket{\psi_t}$ evolves as
\begin{equation}\label{eq:quantum_traj}
d\ket{\psi_t}
=
\left[
-i\hat{H}_{\mathrm{eff}}\,dt
+\sum_i
\left(
\frac{\hat{c}_i}{\sqrt{\langle \hat{c}_i^\dagger \hat{c}_i\rangle_t}}-1
\right)dN_i(t)
\right]\ket{\psi_t},
\end{equation}
with \(\hat{c}_i\) given in Eq.~\eqref{eq:collapse_operators}. Here, \(\langle \hat c_i^\dagger \hat c_i\rangle_{\psi(t)}\) denotes the expectation value taken in the instantaneous trajectory state \(|\psi(t)\rangle\) and
\begin{equation}
\hat{H}_{\mathrm{eff}}
=
\hat{H}-\frac{i}{2}\sum_i \hat{c}_i^\dagger \hat{c}_i,
\end{equation}
Here, $\hat H$ is the system Hamiltonian, which generates the coherent part of the evolution, while the operators $\hat c_i$ describe the dissipative measurement channels associated with the environment. Between detection events, the state evolves under the non-Hermitian effective Hamiltonian $\hat H_{\mathrm{eff}}$, whose imaginary part continuously reduces the norm and encodes the possibility that a jump may occur. The stochastic increments $dN_i(t)\in\{0,1\}$ are Poisson-distributed variables indicating whether a jump occurs in channel $i$ during the interval $dt$. Their mean value, $\mathbb E[dN_i(t)] = \langle \hat c_i^\dagger \hat c_i\rangle_t dt$, gives the jump probability in that interval. When a jump occurs, the state is updated by the normalized action of $\hat c_i$, whereas if no jump occurs the evolution is solely governed by $\hat H_{\mathrm{eff}}$. In this way, the trajectory consists of smooth nonunitary evolution interrupted by random quantum jumps.

At the level of phase-space coordinates, $\boldsymbol r_t=(u_t,v_t)^T$, Eq.~\eqref{eq:quantum_traj} induces a stochastic evolution for each quantum realization of the form
\begin{equation}
d\boldsymbol r_t=\mathbf f(\boldsymbol r_t; \boldsymbol{\eta})\,dt+d\boldsymbol{\xi}_t,
\end{equation}
where $\mathbf f(\boldsymbol r; \boldsymbol{\eta})$ is the same semiclassical flow field appearing in the Wigner evolution, and $d\boldsymbol{\xi}_t$ represents the additional noise amplitude induced by the random occurrence of quantum jumps. In practice, we generate these stochastic trajectories by numerically solving Eq.~\eqref{eq:quantum_traj} with the Monte Carlo wave-function solver implemented in \texttt{QuantumToolbox.jl}~\cite{QuantumToolbox.jl2025}.

\begin{figure*}[!t]
    \centering
    \includegraphics[width=\textwidth]{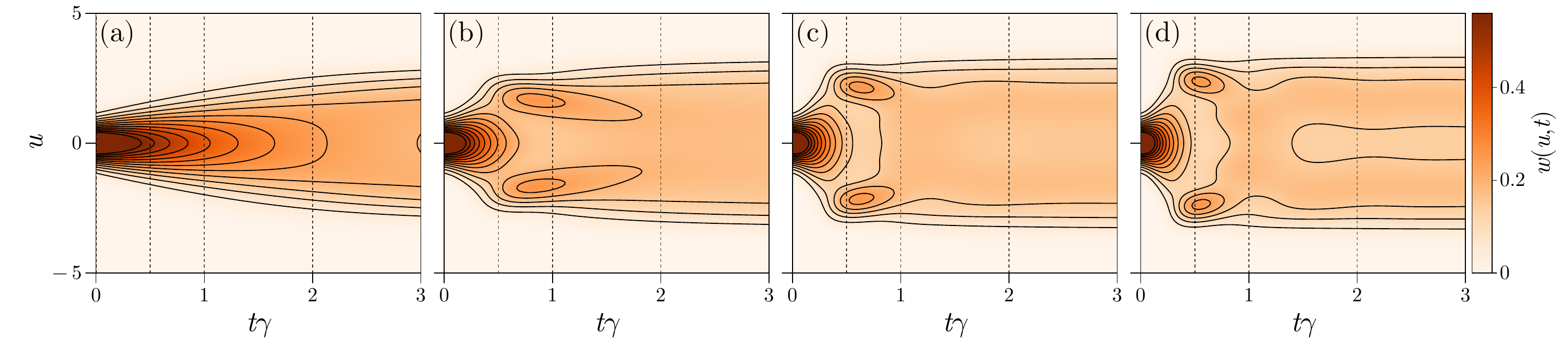}
    \caption{Wigner marginal over the $v$ direction. Same parameters as in Fig.~\ref{fig:6}.}
    \label{fig:S5}
\end{figure*}

To assess how well the classical basin structure survives, we compare the late-time outcome of these stochastic trajectories with the classical basin from which they were initialized. Let $\{B_\alpha\}$ denote the classical basins of attraction obtained from the semiclassical flow. We initialize ensembles of quantum trajectories inside each basin $B_\alpha$ and record the late-time label $\beta$ reached by each realization. This defines the confusion matrix
\begin{equation}
C_{\beta\alpha}
=
\frac{N_{\beta\leftarrow\alpha}}{\sum_{\beta'}N_{\beta'\leftarrow\alpha}},
\end{equation}
where $N_{\beta\leftarrow\alpha}$ is the number of trajectories started in basin $\alpha$ and ending in outcome $\beta$.

In the semiclassical regime one expects $C_{\beta\alpha}\approx \delta_{\beta\alpha}$, while off-diagonal elements quantify fluctuation-induced mixing between distinct basins. 

Figure~\ref{fig:S4}(a) shows $C_{ji}$ for three representative multistable phases. We also include a simple scalar measure of this mixing, $\epsilon = 1-(\sum_\alpha C_{\alpha\alpha})/N_B$, with $N_B$ the number of classical basins. Small $\varepsilon$ means weak basin mixing and indicates that the molecule remains an effective scaffold to the quantum dynamics, while $\varepsilon=O(1)$ signals strong mixing and a gradual loss of sharp classical connectivity. The left column corresponds to a regime in which trajectories initialized in the LC basin remain strongly confined, while trajectories in the basins of the FPs exhibit leakage toward the LC. In the middle column, after the homoclinic bifurcation, the pattern of basin mixing changes consistently with the altered global connectivity of the flow. The system displays a more robust two-basin structure, with significantly reduced interbasin mixing.

These results support the interpretation of the molecule as a meaningful description of the dynamics beyond the strict classical limit, although its predictive power gradually diminishes as fluctuations are introduced into the system. 

\section{Alternative experimentally accessible observables}\label{app:wigner_marginals}

To gain further sensitivity to the changes in relaxation pathways induced by the homoclinic (global) transition in Fig.~\ref{fig:3}(a), one can turn to phase-space-sensitive observables that are simpler than full Wigner tomography, such as quadrature marginals obtained from repeated homodyne or heterodyne measurements. These measurements directly sample a chosen field quadrature, e.g. via balanced homodyne detection~\cite{leonhardt1997measuring}, allowing one to reconstruct its probability distribution without performing the full parity-based tomography required for the complete Wigner function.

In the present problem, due to the $\mathbb{Z}_2$ symmetry and our choice of two-photon driving phase $\theta=0$, the relevant phase-space structures are organized predominantly along the $u$ axis. Therefore, a natural choice is to consider the marginal over the $v$ direction,
\begin{equation}
    w(u,t)=\int dv\, W(u,v;t),
\end{equation}
where $W(u,v;t)$ is the Wigner distribution at time $t$. The marginal $w(u,t)$ still retains partial information about whether probability is concentrated near central regions, expelled from a forbidden annulus, or split between different basins.

Figure~\ref{fig:S5} shows that the quadrature marginal still tracks how probability is redistributed along the symmetry axis selected by the FP structure. In panel (a), the weight remains centered and broadens smoothly, consistent with direct relaxation toward the LC. In panels (b) and (c), pronounced off-center lobes develop, signaling transient occupation of the symmetry-related FP regions. In panel (d), the balance between central and off-center weight is modified again, revealing the change in unstable phase-space organization across the global transition. In particular, this observable still reveals partial signatures of the flow-topology transition, including the difference between the phases of Fig.~\ref{fig:3}(d) and Fig.~\ref{fig:3}(e) that is only weakly visible in the spectrum.

\makeatletter
\immediate\write\@auxout{\string\citation{apsrev42Control}}
\makeatother

\end{document}